\documentclass[]{article}
\usepackage[utf8]{inputenc}
\usepackage{amsmath}
\usepackage{amssymb}
\usepackage{graphicx}
\usepackage{authblk}
\usepackage{ulem}

\title{\textbf{Motility-Induced Clustering and Meso-Scale Turbulence in Active Polar Fluids}}
\author[1]{Vasco M. Worlitzer}
\author[2]{Gil Ariel}
\author[3,4]{Avraham Be'er}
\author[5]{Holger Stark}
\author[1]{Markus Bär}
\author[1]{Sebastian Heidenreich}
\affil[1]{Department of Mathematical Modelling and Data Analysis, Physikalisch-Technische Bundesanstalt Braunschweig und Berlin, Abbestrasse 2-12, D-10587 Berlin, Germany}
\affil[2]{Department of Mathematics, Bar-Ilan University, Ramat Gan 52000, Israel}
\affil[3]{Zuckerberg Institute for Water Research, The Jacob Blaustein Institutes for Desert Research, Ben-Gurion University of the Negev, Sede Boqer Campus 84900, Midreshet Ben-Gurion, Israel}
\affil[4]{Department of Physics, Ben-Gurion University of the Negev, Beer Sheva 84105, Israel}
\affil[5]{Institute of Theoretical Physics, Technische Universität Berlin, Hardenbergstrasse 36, D-10623 Berlin, Germany}

\begin{document}

\maketitle

\begin{abstract}
    Meso-scale turbulence was originally observed experimentally in various suspensions of swimming bacteria, as well as in the collective motion of active colloids. The corresponding large scale dynamical patterns were reproduced in a simple model of a polar fluid, assuming a constant density of active particles. Recent, more detailed studies in a variety of experimental realizations of active polar fluids revealed additional interesting aspects, such as anomalous velocity statistics and clustering phenomena. Those phenomena cannot be explained by currently available models for active polar fluids. Herein, we extend the continuum model suggested by Dunkel et al.\ to include density variations and a local feedback between the local density and self-propulsion speed of the active polar particles. If the velocity decreases strong enough with the density, a linear stability analysis of the resulting model shows that, in addition to the short-wavelength instability of the original model, a long-wavelength instability occurs. This is typically observed for high densities of polar active particles and is analogous to the well-known phenomenon of motility-induced phase separation (MIPS) in scalar active matter. We determine a simple phase diagram indicating the linear instabilities and perform systematic numerical simulations for the various regions in the corresponding parameter space. The interplay between the well understood short-range instability (leading to meso-scale turbulence) and the long-range instability (associated with MIPS) leads to interesting dynamics and novel phenomena concerning nucleation and coarsening processes. Our simulation results display a rich variety of novel patterns, including phase separation into domains with dynamically changing irregularly shaped boundaries. Anomalous velocity statistics are observed in all phases where the system segregates into regions of high and low densities. This offers a simple explanation for their occurrence in recent experiments with bacterial suspensions.
\end{abstract}

\section{Introduction}

Exploring active matter has become a popular subject in contemporary physics leading to many new insights into a large variety of intriguing systems. Active systems are ubiquitous in nature, thereby drawing interest from different scientific communities (physics, chemistry, biology, material science, ecology, robotics) and offering a wealth of surprising dynamic phenomena \cite{Ramaswamy2010, Vicsek2012, Romanczuk2012b, Marchetti2013, Bechinger2016, Beer2019}. Moreover, they provide many new challenges for our understanding of non-equilibrium systems. Realizations of active matter systems range from intracellular processes and bacterial suspensions \cite{Ramaswamy2010, Beer2019, Zhang2010, Koch2011}, artificial Janus particles \cite{Walther2013,Nishiguchi2015,Buttinoni2013} to schools of fish and flocks of birds \cite{Katz2011,Cavagna2010}. More recent reviews have focused on the prominent role of models with alignment interaction \cite{Chate2020} and on anisotropic, self-propelled particles \cite{Bar2020} as well as on the large variety of computational approaches to active matter \cite{Shaebani2020} and on a roadmap outlining a multitude of promising directions for the field \cite{Gompper2020}.

Among various collective states that characterize active matter, one phenomenon of general interest is meso-scale turbulence. Meso-scale turbulence is reported in various experimental studies, for example for suspensions of \textit{Bacillus subtilis} \cite{Dombrowski2004,Cisneros2007}, \textit{Escherichia coli} \cite{Ishikawa2011,Liu2012} and \textit{Serratia marcescens} \cite{Steager2008}. The main feature (and difference to ordinary inertial turbulence) of meso-scale turbulence in bacterial suspensions is the appearance of a characteristic length scale \cite{Dombrowski2004,Ishikawa2011,Wensink2012a,Bratanov2015,Zhang2009,James2018,James2018b}. A continuum model that agrees with experimental findings for wild-type \textit{Bacillus subtilis} suspensions was presented in \cite{Wensink2012a,Dunkel2013a,Dunkel2013b,Heidenreich2016,Reinken2018a}. When solving this model numerically for a broad range of parameters, the observed velocity statistic is close to Gaussian \cite{Ariel2018} in agreement with experimental findings \cite{Wensink2012a}. 

Recent work on discrete models with self-propelled rods revealed that effective polar alignment is often observed in self-propelled rods with steric repulsion \cite{Grossmann2020, Jayaram2020}. It is important to note that the model for meso-scale turbulence analyzed in this paper is appropriate for the description of polar fluids and does not apply to active turbulence in so called “active nematics”, see e.\ g.\ \cite{Doostmohammadi2018, Alert2020}, wherein a different kind of turbulence without a characteristic length-scale, compare discussion in \cite{Bar2020}, is observed. Current experiments on engineering of vortex lattices in bacterial suspensions \cite{Nishiguchi2018} indeed showed that a typical length-scale is controlling the behavior in meso-scale turbulence. As a result, the continuum model of meso-scale turbulence described in detail above allowed to reproduce the characteristics of these experiments \cite{Reinken2020}.

However, in recent experiments on \textit{Bacillus subtilis} suspensions anomalous velocity statistics have been observed. For example, swarms of very short or very long cells (compared to the wild-type) show anomalous velocity statistics \cite{Ilkanaiv2017}. Moreover, adding sublethal concentrations of antibiotics to wild-type swarms produces anomalous velocity statistics \cite{Benisty2015}. The deviations from normal statistics are quantified by measuring the kurtosis $\kappa$ (scaled fourth moment), where $\kappa\neq3$ indicates anomalous statistics. Such anomalous statistics are not reported for the theory presented in \cite{Wensink2012a,Dunkel2013b}. Hence, a theory accounting for anomalous statistics in meso-scale turbulent systems is still lacking.

We present and analyse a minimal model based on \cite{Wensink2012a,Dunkel2013b}, exhibiting meso-scale turbulence and anomalous statistics. The main idea is summarized as follows: The model introduced in \cite{Wensink2012a,Dunkel2013b} assumes constant density and constant self-propulsion speed. We relax these assumptions by allowing for velocity variations mediated by density variations. This is a very natural assumption, as a dependency of the speed on the density is reported for \textit{Bacillus subtilis} suspensions in several experimental studies \cite{Ariel2018,Sokolov2007}. More specifically, we borrow ideas from motility-induced phase separation (MIPS) \cite{Cates2010,Fily2012,Bialke2013} to model density variations.

Combining ideas from meso-scale turbulence and MIPS is intriguing from a general point of view as well. In bacterial suspensions steric interactions (through volume exclusion), alignment (through elongated shapes) and hydrodynamic interactions (through self-propulsion in the surrounding medium) are assumed to be present simultaneously. Considering steric interactions and alignment individually gives rise to MIPS \cite{Cates2010,Fily2012,Bialke2013} and global order in Vicsek-like models \cite{Vicsek1995,Toner1995,Toner1998,Toner2005} respectively, while meso-scale turbulence results from the combination of alignment and hydrodynamics \cite{Wensink2012a,Dunkel2013b,Heidenreich2016,Reinken2018a}. However, a theory aiming to comprehensively describe the dynamics of bacterial suspensions needs to incorporate all three of these effects and the interplay between them. Recently, several studies focusing on the interplay of steric interactions and alignment in active matter \cite{VanDamme2019,Grossmann2020,Jayaram2020,Sese-Sansa2018,Shi2018,Geyer2019,Barre2014,Theers2018,VanDerLinden2019} report interesting and sometimes contradicting results. However, hydrodynamic interactions are commonly neglected. Our model features elements from all three of these prominent theories of active matter (MIPS, global order, meso-scale turbulence) and connects them in a minimalist fashion. Hence, our work contributes to the current discussion on how to connect different branches of active matter and provides an insight into the expected dynamics. We remark that in this study we mainly focus on the interplay between MIPS and the finite-wavelength instability arising due to hydrodynamic interactions, while global order will be of less relevance.

The structure of the paper is as follows: In the second section we propose a phenomenological model based on continuum models well established in the literature, which combines their central features. In the third section we present a linear stability analysis, hinting at the expected dynamics. In the fourth section we present numerical solutions of our model, sketch a phase portrait and discuss the observed anomalous velocity statistics.

\section{Modeling Approach}\label{sec:Modeling}

We present our phenomenological model in the following steps: First, we revisit two continuum models. The first model describes MIPS, while the second model describes meso-scale turbulence. Based on these models, we propose a minimal phenomenological model that combines their main features.

When discussing continuum models of active matter, it is helpful to keep the microscopic picture in mind. We consider active particles that self-propel with some speed along their individual axes. Coarse-graining gives an averaged polarization, which, multiplied with the speed, coincides with the macroscopic velocity. Hence, the polarization plays a dual role of both order parameter and velocity field \cite{Ramaswamy2010,Marchetti2013,Fodor2018a}.

\subsection{Revisiting established models}\label{sec:BaseModels}
A minimal hydrodynamic model to describe MIPS was presented in \cite{Fily2012,Bialke2013}. It consists of coupled equations for the particle density $\rho$ and polarization density field $\mathbf{p}$
\begin{subequations}\label{eq:MIPS}
    \begin{align}\label{eq:MIPS_Continuity}
        \partial_t\rho &= -\nabla \cdot [v(\rho)\mathbf{p}] + D\Delta\rho,\\
        \label{eq:MIPS_Pol}
        \partial_t\mathbf{p} &= -\frac{1}{2}\nabla [v(\rho)\rho] + D\Delta\mathbf{p} - D_r\mathbf{p},
    \end{align}
\end{subequations}
where $D$ and $D_r$ are diffusion coefficients. The polarization density ${\bf p}$ is given by the averaged orientation of the self-propelled particles. The coupling between the density $\rho$ and the polarization density $\mathbf{p}$ is achieved through a density-dependent speed $v(\rho)$, with
\begin{equation}\label{eq:v_rho_linear}
    v(\rho) = v_0 - \zeta\rho
\end{equation}
modelling a decrease of the self-propulsion speed at very high density (jamming) \cite{Ariel2018,Beer2020}. A transition from a homogeneous density profile to phase separation is encountered for sufficiently high densities (or alternatively a strong enough damping constant $\zeta$). The phase separation can be understood by assuming a slow variation of $\mathbf{p}$ in time and space. Setting the corresponding derivatives in Eq.\ \eqref{eq:MIPS_Pol} to zero and substituting the resulting expression for $\mathbf{p}$ into Eq.\ \eqref{eq:MIPS_Continuity} gives
\begin{equation}\label{eq:MIPS_effective}
    \partial_t\rho = \nabla\cdot\mathcal{D}(\rho)\nabla\rho,\quad \mathcal{D}(\rho) = D+\frac{v(\rho)\left[v(\rho)+v'(\rho)\rho\right]}{2D_r}.
\end{equation}
Above a critical density, the sign of the effective diffusion coefficient $\mathcal{D}(\rho)$ changes due to $v'(\rho)<0 $, triggering phase separation. For more details refer to \cite{Fily2012,Bialke2013,Fodor2018a}.

Taking a different approach, in \cite{Wensink2012a,Dunkel2013b} the authors present a model which reproduces the statistical features of meso-scale turbulence as observed in dense suspensions of \textit{Bacillus subtilis}. This model was proposed on a phenomenological basis and was later derived from a microscopic microswimmer model \cite{Heidenreich2016,Reinken2018a}. It can be regarded as the combination of a (simplified) Toner-Tu model \cite{Toner1995,Toner1998} and a fourth-order term as in the Swift-Hohenberg equation \cite{Swift1977}. The time-dependent polarization density evolves according to
\begin{subequations}\label{eq:Incomp}
	\begin{align}\label{eq:pIncomp}
	&\partial_t \mathbf{p} +\lambda_0 (\mathbf{p}\cdot\nabla)\mathbf{p}
	=-\nabla P -(A+C|\mathbf{p}|^2)\mathbf{p}+\Gamma_0\Delta\mathbf{p}-\Gamma_2\Delta^2\mathbf{p},\\
	\label{eq:divFree}
	&\nabla\cdot\mathbf{p}=0.
	\end{align}
\end{subequations}
The density is assumed to be constant, which leads to the incompressibility condition Eq.\ \eqref{eq:divFree} and introduces the Lagrange multiplier $P$ enforcing this condition. Rewriting model \eqref{eq:Incomp} in potential form
\begin{equation}\label{eq:PotentialForm}
    \partial_t \mathbf{p} +\lambda_0 (\mathbf{p}\cdot\nabla)\mathbf{p} =-\frac{\delta\mathcal{F}}{\delta\mathbf{p}}, \quad\nabla\cdot\mathbf{p}=0,
\end{equation}
with
\begin{equation}
    \mathcal{F} = P(\nabla\cdot\mathbf{p}) + \frac{1}{2}A|\mathbf{p}|^2 + \frac{1}{4}C|\mathbf{p}|^4 + \frac{1}{2}\Gamma_0(\nabla\mathbf{p})^2 + \frac{1}{2}\Gamma_2(\nabla\nabla\mathbf{p})^2
\end{equation}
shows that the dynamics of $\mathbf{p}$ is governed by pure relaxational dynamics derived from $\mathcal{F}$ and a convective part $\lambda_0 (\mathbf{p}\cdot\nabla)\mathbf{p}$. For the bulk terms in $\mathcal{F}$ we distinguish between two regimes: For $A>0$ and $C>0$ these terms stabilize a disordered state. For $A<0$ and $C>0$ they represent a double-well potential, forcing a nonzero magnitude of the polarization density. Physically, the rotational symmetry is broken spontaneously and a globally ordered state with $|\mathbf{p}|=\sqrt{-A/C}$ is stable in this case. Hence, for $\Gamma_0>0$ and $\Gamma_2=0$ the model \eqref{eq:Incomp} reduces to the Toner-Tu theory. However, if $\Gamma_0<0$ and $\Gamma_2>0$ a finite-wavelength instability is introduced, similar as in the Swift-Hohenberg equation \cite{Swift1977}. This instability destabilizes both the disordered and globally ordered state simultaneously, leading to meso-scale turbulence. Setting $\Gamma_0<0$ is justified by physical arguments. Indeed, model \eqref{eq:Incomp} can be derived from microscopic considerations using a coarse-graining procedure \cite{Heidenreich2016,Reinken2018a}, which indeed leads to $\Gamma_0<0$ due to activity and hydrodynamics. The competition between alignment and hydrodynamics sets the effective length scale of the evolving pattern. In this derivation it also becomes apparent that the coefficients in model \eqref{eq:Incomp} should, in general, also depend on density.

\subsection{Extended model}
Two major assumptions underlie model \eqref{eq:Incomp}: That density is constant and that the particles propel with constant speed $v_0$ along their orientation. We relax these assumptions and replace them by expressions motivated by model \eqref{eq:MIPS}.

First, we assume that the velocity is density-dependent, i.e. we replace the constant speed $v_0$ with a density-dependent speed $v(\rho)$. Such an assumption is very natural in realistic active matter systems as excluded volume as well as collective effects lead to a density-dependent speed \cite{Buttinoni2013,Ariel2018,Sokolov2007,Fily2012}. Next, we have to replace the incompressibility condition by an evolution equation for the density. A natural choice is a continuity equation consisting of the divergence of the mass flux and a diffusive term
\begin{equation}\label{eq:continuity}
	\partial_t \rho = -\nabla\cdot[v(\rho)\mathbf{p}]+D\Delta\rho.
\end{equation}
Moreover, this equation agrees with the ones derived by \cite{Fily2012,Bialke2013} for MIPS (see Eq.\ \eqref{eq:MIPS_Continuity}) as well as the one derived for model \eqref{eq:Incomp} (in a suitable defined limit).

In model \eqref{eq:Incomp} the coupling to the (degenerate) density equation is accomplished via the Lagrange multiplier $P$ acting as a pressure. As we replace the incompressibility condition Eq.\ \eqref{eq:divFree} with the continuity equation Eq.\ \eqref{eq:continuity} an explicit coupling in terms of the density is needed. We choose 
\begin{equation}\label{eq:pressure}
	P(\rho) = \frac{1}{2}v(\rho)\rho.
\end{equation}
Such a term appears naturally when coarse-graining microscopic models that incorporate self-propulsion. Indeed, this term is reported for all comparable systems we are aware of \cite{Fily2012,Bialke2013,Grossmann2020,Jayaram2020,Geyer2019,Speck2014,Speck2015}, see also Eq.\ \eqref{eq:MIPS_Pol}. While the details of the underlying microscopic model and coarse-graining procedure (especially the choice of an appropriate closure) might introduce additional terms to the dynamics of the polarization density $\mathbf{p}$, a term as in Eq.\ \eqref{eq:pressure} will always be present. Secondly, one can think of Eq.\ \eqref{eq:pressure} as a low order approximation of the pressure, disregarding higher order coupling between $\rho$ and $\mathbf{p}$. We note that Eq.\ \eqref{eq:pressure} can only formally be regarded as a pressure. Determining the pressure of active fluids is in general a complicated task \cite{Takatori2014,Solon2015a,Solon2015c}, especially when accounting for hydrodynamic interactions. Alternatively, a virial expansion \cite{Falasco2016} or a treatment as in the Toner-Tu theory \cite{Toner1995, Toner1998} is possible.

As briefly mentioned in section \ref{sec:BaseModels}, all coefficients of model \eqref{eq:Incomp} are, in principle, dependent on the density. An appropriate rescaling reduces the possible density-dependent parameters to $\lambda_0,\Gamma_0$ and $A$. Experimental and numerical findings agree that the characteristic length scale set by $\Gamma_0$ in the turbulent regime does not depend on the (overall) density \cite{Sokolov2012}. Hence, we drop that dependency. Furthermore, for simplicity we assume $\lambda_0$ to be independent of density as well. Note though that earlier studies suggest a non-monotone dependence of $\lambda_0$ on $\rho$ \cite{Heidenreich2016}. This leaves $A$ as the only density-dependent parameter. In fact, the transition from a dilute, disordered state to a dense, globally ordered state in the Vicsek model can be explained by a change of sign of $A$ through an increased density. Altogether, our phenomenological model in its most general form is given by
\begin{subequations}\label{eq:Dynamics}
	\begin{align}
	\partial_t \rho &= -\nabla\cdot[v(\rho)\mathbf{p}]+D\Delta\rho,\\
	\partial_t \mathbf{p} +\lambda_0 (\mathbf{p}\cdot\nabla)\mathbf{p}
	&=-\frac{1}{2}\nabla [v(\rho)\rho] - \left[A(\rho)+C|\mathbf{p}|^2\right]\mathbf{p} +\Gamma_0\Delta\mathbf{p}-\Gamma_2\Delta^2\mathbf{p}.
	\end{align}
\end{subequations}
Overall, our model is essentially a minimal model that incorporates the main features of the models presented in the previous section. Moreover, the three instabilities can be tuned independently through the coupling terms and $\Gamma_0$.

\section{Stability Analysis}\label{sec:stability}
As a first insight into the dynamics expected from model \eqref{eq:Dynamics} we perform a linear stability analysis for the steady states of the model. Clearly, a trivial steady state is given by $(\rho,\mathbf{p})=(\rho_0,\mathbf{p}_0)$, where $\rho_0$ and $\mathbf{p}_0$ are uniform in space. In the following we distinguish between the case $\mathbf{p}_0=0$ (disorder) and $|\mathbf{p}_0|>0$ (global polar order). Due to the special structure of the stability matrix (see Appendix \ref{app:stability} for details), the dispersion relations for the disordered state can be computed analytically. This is also possible for the polar state. However, finding the dispersion relation, i.e.\ solving for the eigenvalues of the stability matrix, produces lengthy expressions offering little insight. Hence, we only present numerical results for the polar state.

\subsection{Disordered State}\label{sec:stab_dis}
Linearizing Eq.\ \eqref{eq:Dynamics} around the steady state $(\rho,\mathbf{p}) \equiv (\rho_0,0)$ and expanding perturbations into Fourier modes reveals the dispersion relations
\begin{subequations}\label{eq:Dispersion}
	\begin{align}
	\label{eq:sigma1}
	\sigma_1(k) &= -A(\rho_0)-\Gamma_0 k^2-\Gamma_2 k^4,\\
	\label{eq:sigma2}
	\sigma_{2,3}(k) &= \frac{1}{2}\left[-A(\rho_0)-(\Gamma_0+D)k^2-\Gamma_2k^4
	\pm \sqrt{r(k)}\right],
	\end{align}
\end{subequations}
where $k=|\mathbf{k}|$ is the magnitude of the wavevector $\mathbf{k}=(k_1,k_2)$. See Appendix \ref{app:stability} for details. Furthermore, we introduced 
\begin{equation}
    \begin{aligned}
        r(k) =& A(\rho_0)^2+\left[2A(\rho_0)(\Gamma_0-D)-4\gamma\right]k^2
	    +\left[2A(\rho_0)\Gamma_2+(\Gamma_0-D)^2\right]k^4\\
	    &+2(\Gamma_0-D)\Gamma_2 k^6
	    +\Gamma_2^2 k^8,
    \end{aligned}
\end{equation}
where the constant $\gamma$, quantifying the coupling to the density-dependent propulsion speed, is given by
\begin{equation}\label{eq:coupling_combined}
    \gamma = \frac{1}{2}v(\rho_0)\left[v(\rho_0)+v'(\rho_0)\rho_0\right].
\end{equation}

\begin{figure}[!htb]
	\centering
	\includegraphics[width=\linewidth, height=\textheight,keepaspectratio]{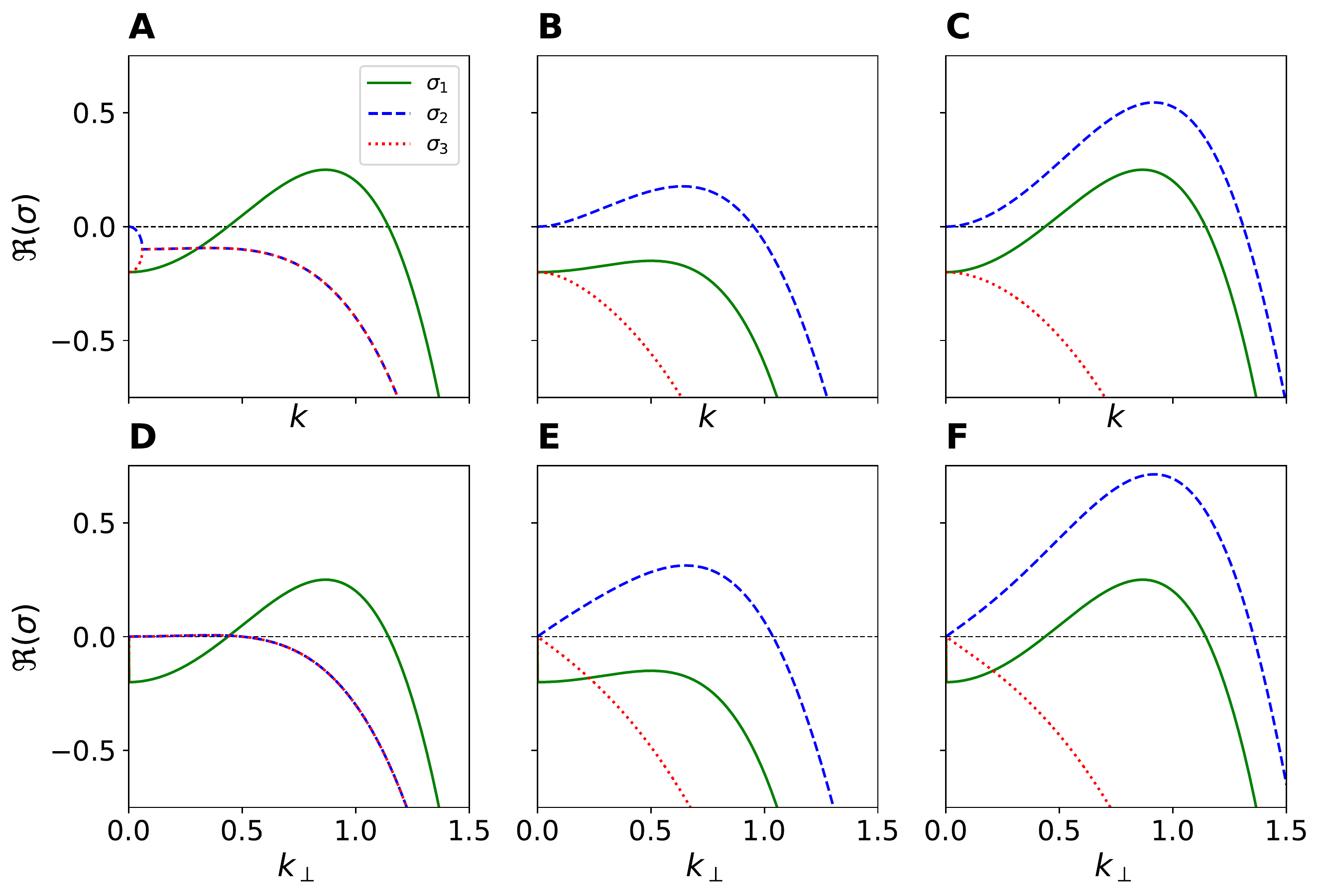}
	\caption{Dispersion relations for the disordered state (\textbf{A}-\textbf{C}) and the polar state in the direction $\mathbf{k}_\perp$ perpendicular to $\mathbf{p}_0$ (\textbf{D}-\textbf{F}). For \textbf{A},\textbf{D} we set $\Gamma_0<\Gamma_0^c$ and $\zeta<\zeta^c$. For \textbf{B},\textbf{E} we set $\Gamma_0>\Gamma_0^c$ and $\zeta>\zeta^c$ and for \textbf{C},\textbf{F} we set $\Gamma_0<\Gamma_0^c$ and $\zeta>\zeta^c$. In all subplots $v(\rho)$ is chosen as Eq.\ \eqref{eq:v_rho_linear}.}
	\label{fig:Disp}
\end{figure}

The first eigenvalue $\sigma_1(k)$ does not contain any coupling terms nor any contributions from the density equation. Furthermore, the corresponding eigenvector is given by
\begin{equation}
	\mathbf{v}_1 = (0,-k_2,k_1),
\end{equation}
which is independent of density. Hence, $\sigma_1(k)$ solely affects the stability of the polarization density $\mathbf{p}$. Moreover, the same dispersion relation is found in model \eqref{eq:Incomp}, see \cite{Wensink2012a,Dunkel2013b}. Therefore, model \eqref{eq:Dynamics} inherits the finite-wavelength instability of the polarization for sufficiently small $\Gamma_0<0$ from model \eqref{eq:Incomp}. This instability is characterized by a band of unstable modes bounded away from zero as can be seen in figure \ref{fig:Disp}\textbf{A}. It is straightforward to compute the critical parameter from Eq.\ \eqref{eq:sigma1} as
\begin{equation}\label{eq:Gamm0crit}
	\Gamma_0^c = -\sqrt{4A(\rho_0)\Gamma_2}.
\end{equation}

The qualitative behavior of the other eigenvalues $\sigma_{2,3}(k)$ cannot be read off directly from Eq.\ \eqref{eq:sigma2}. Instead, performing a small wavenumber expansion 
\begin{equation}\label{eq:sigma2approx}
	\sigma_2(k) \approx -\left(D + \frac{\gamma}{A(\rho_0)}\right)k^2 = -\mathcal{D}(\rho_0)k^2
\end{equation}
reveals a long-wavelength instability for $\mathcal{D}(\rho_0)<0$. From this expression and Eq.\ \eqref{eq:coupling_combined} we can calculate critical parameters by specifying $v(\rho)$. The resulting long-wavelength instability is pictured in figure \ref{fig:Disp}\textbf{B}. As expected from our modeling approach, this instability is similar to the one reported for MIPS, see section \ref{sec:BaseModels} and \cite{Fily2012,Speck2015}. Note that $\mathcal{D}(\rho_0)$ coincides with $\mathcal{D}(\rho)$ introduced in Eq.\ \eqref{eq:MIPS_effective} for the choice $\rho=\rho_0$.
 
While both instabilities have been studied in great detail separately, to the authors knowledge, a situation as depicted in figure \ref{fig:Disp}\textbf{C}, where both instabilities are present at the same time, has not been studied yet. As we will show in the next section, this leads to interesting dynamics.

\subsection{Polar State}
There is an additional long-wavelength instability of the disordered state for $A(\rho_0)<0$. As discussed in section \ref{sec:BaseModels}, a steady state with polar order, i.e.\ $(\rho,\mathbf{p}) \equiv (\rho_0,\mathbf{p}_0)$ and $|\mathbf{p}_0| = \sqrt{-A(\rho_0)/C}$, emerges in this situation. The stability analysis of the polar state can be carried out similar to the disordered case. However, the resulting dispersion relations are lengthy and intricate, hampering an intuitive interpretation. Numerical computation of the eigenvalues reveals instabilities similar to the disordered state. There is a finite-wavelength instability for $\Gamma_0\leq\Gamma_0^c$ perpendicular to $\mathbf{p}_0$ (see figure \ref{fig:Disp}\textbf{D}) and a long-wavelength instability above a critical coupling constant, also perpendicular to $\mathbf{p}_0$ (see figure \ref{fig:Disp}\textbf{E}). Finally, both instabilities can be present as in the disordered case, see figure \ref{fig:Disp}\textbf{F}. Interestingly, the numerically computed critical values for both instabilities are the same as in the disordered case. Hence, the disordered and polar states loose stability simultaneously, indicating the existence of a new dynamical attractor.

\section{Numerical solution of the model equations}\label{sec:pp}
We now explore the dynamics produced by model \eqref{eq:Dynamics} numerically. For this purpose, we have to specify the coupling terms $v(\rho)$ and $A(\rho)$. As the model is complex, we aim for simple coupling terms in order to ease the numerical burdens and reduce the amount of possible parameters. Hence, we choose $v(\rho) = v_0 - \zeta\rho$, i.e.\ we study the linear case of Eq.\ \eqref{eq:v_rho_linear}. A monotone decrease with density is motivated by crowding effects, i.e.\ self-propulsion is counteracted by steric hindrance in dense areas. The linear model Eq.\ \eqref{eq:v_rho_linear} with coefficient $\zeta$ was discussed in \cite{Fily2012,Bialke2013,Speck2014,Speck2015} and derived as a first-order approximation from microscopic coniderations. Other monotonically decreasing functions of density have been studied in the literature as well (see \cite{Cates2010} and \cite{Geyer2019} for an exponential or hyperbolic tangent dependence respectively).

In addition, to further simplify the analysis, we choose $A(\rho) \equiv A > 0$. While such a choice might seem arbitrary at first glance, complex non-equilibrium dynamics can be observed for sufficiently small $\Gamma_0$, even for $A>0$, due to local shear stresses. A profound analysis on the influence of the potential terms with coefficients $A$ and $C$ on the dynamics of model \eqref{eq:Incomp} can be found in \cite{Dunkel2013b}. Therein, the authors conclude that the main difference is an absence of jets for $A>0$. From a general point of view, the choice $A(\rho) \equiv A > 0$ disregards the polar state and its effects on the dynamics, which allows us to focus on the interplay between meso-scale turbulence and MIPS and to reduce the amount of parameters by setting $C=0$.

First, we will present a numerical phase portrait. As the model includes several parameters, we have to restrict ourself to a low-dimensional cut in parameter space. We study the model in the space spanned by $\Gamma_0$ and $\zeta$. These parameters can be used to control the instabilities independently. Furthermore, we can compare critical values computed numerically with the ones found from the stability analysis in section \ref{sec:stab_dis}. The phenomenological parameters can be related to physical properties by examining microscopic models. While $\Gamma_0$ is determined by the characteristics of the surrounding fluid and the activity (see \cite{Heidenreich2016,Reinken2018a}), the parameter $\zeta$ depends on the details of the repulsive interactions (see \cite{Bialke2013}). We then give a qualitative description of the dynamical phases encountered, when the different instabilities are present. Finally, we discuss the anomalous velocity statistics observed in more detail.

\subsection{Phase Portrait}\label{sec:phase_portrait}
\begin{figure}[!h]
    \centering
    \includegraphics[scale=0.5]{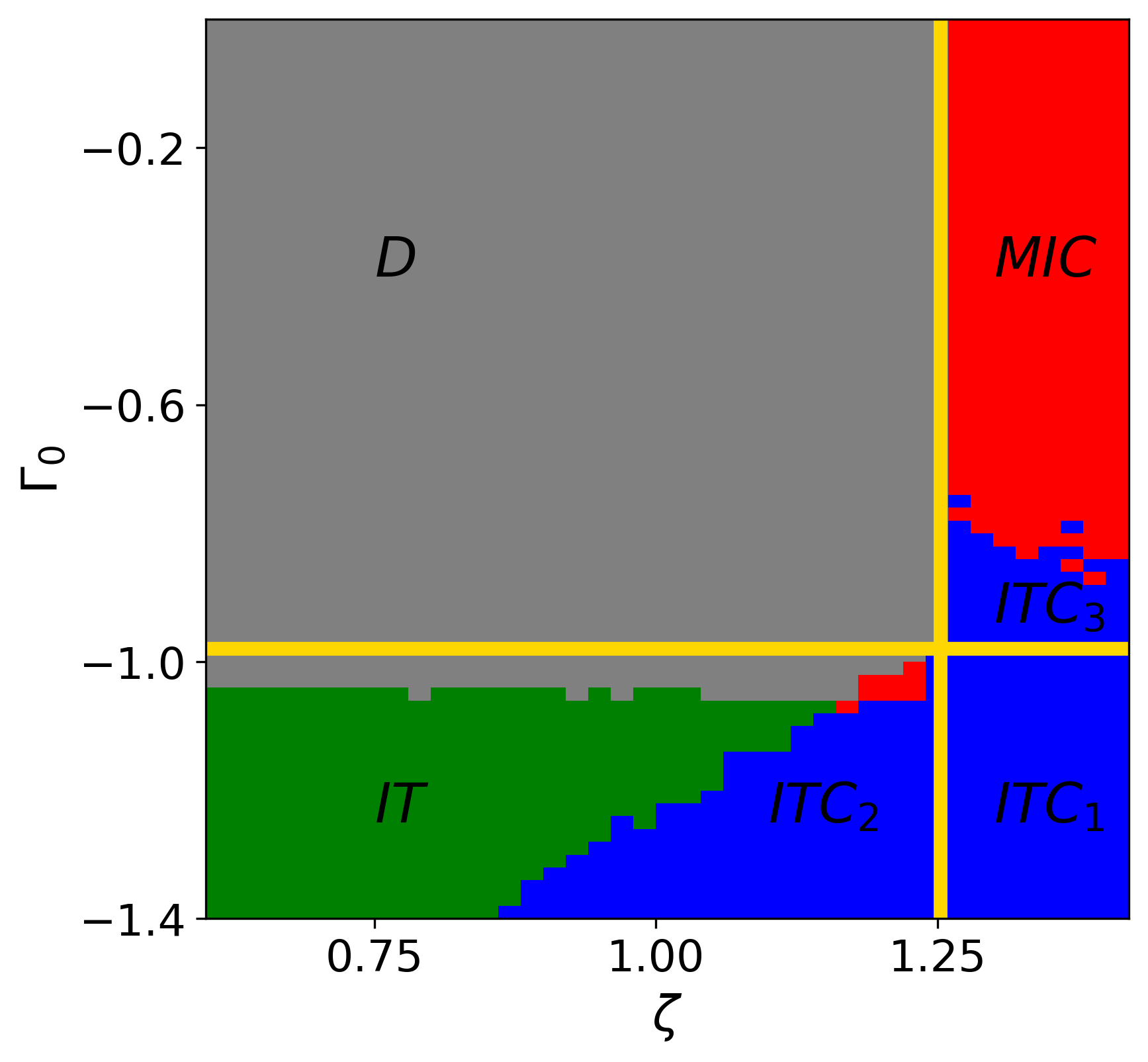}
    \caption{Phase portrait in the $\zeta$-$\Gamma_0$ plane. Phases are color coded, grey = disordered (D), green = Isotropic Turbulence (IT), red = Motility-Induced Clustering (MIC) and blue = Isotropic Turbulence with Clustering (ITC). Additionally, region ITC is subdivided into three subregions. Simulation parameters are listed in table \ref{tab:sim_parameters} in Appendix \ref{app:sim_parameters}.}
	\label{fig:phase_portrait}	
\end{figure}

To obtain a phase portrait, we numerically solve Eq.\ \eqref{eq:Dynamics} for slightly perturbed homogeneous initial conditions. The exact simulation setup can be found in Appendix \ref{app:phase_identifiers}, details on the numerical implementation are provided in Appendix \ref{app:numerics}. To distinguish phases numerically, we introduce two quantifiers: the enstrophy as a measure for the presence of vortices and the modality of the density distribution to detect clustering. Details can be found in Appendix \ref{app:phase_identifiers}. Using these indicators, the phase portrait figure \ref{fig:phase_portrait} is numerically calculated, where the different colors correspond to the phases. Phase boundaries expected from the linear stability analysis are depicted as yellow lines. They can be obtained by calculating critical parameters, which we already determined for $\Gamma_0^c$ in Eq.\ \eqref{eq:Gamm0crit}. Similarly, the critical damping parameter $\zeta^c$ can be obtained by using Eq.\ \eqref{eq:sigma2approx} and  Eq.\ \eqref{eq:v_rho_linear}, giving
\begin{equation}\label{eq:zetacrit}
    \zeta^c = \frac{1}{4\rho_0}\left(3v_0 - \sqrt{v_0^2-16A(\rho_0)D}\right)\approx \frac{v_0}{2\rho_0},
\end{equation}
where the last approximation holds, if the product $A(\rho_0)D$ is small. 


\subsection{Qualitative phase descriptions}\label{sec:pp_simple}

Model \eqref{eq:Dynamics} produces a wealth of new dynamics, which cannot be covered entirely within this work. We therefore only give a qualitative description of the phases characterized by our coarse numerical measures.\\

\begin{figure}[!h]
	\centering
	\includegraphics[width=\linewidth, height=\textheight,keepaspectratio]{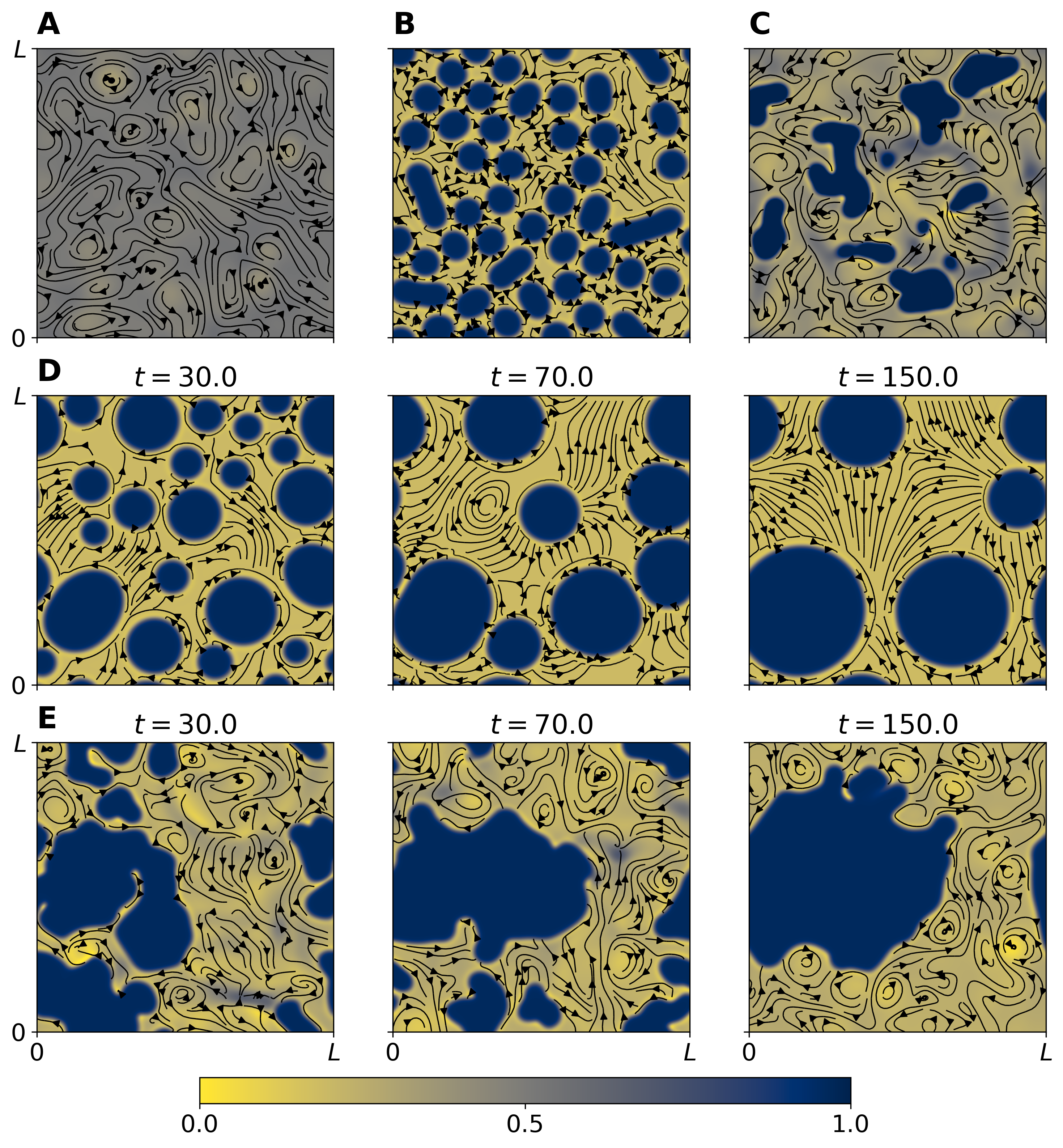}
	\caption{Typical snapshots of the dynamics for different points in the phase portrait, \textbf{A} Isotropic Turbulence \textbf{B}, \textbf{D} Motility-Induced Clustering and Motility-Induced Phase Separation \textbf{C}, \textbf{E} Isotropic Turbulence with Clustering. The rescaled density values are given by the colorbar. Streamlines of the mass flux are only plotted in the dilute phase for better visibility. Snapshots in \textbf{A}-\textbf{C} are taken at $t=30$, showing quasi-static states. In \textbf{D} and \textbf{E} snapshots at times $t=30,70,150$ are shown to visualize long time dynamics without and with turbulence respectively. Simulation parameters are detailed in Appendix \ref{app:sim_parameters}.}
	\label{fig:snapshots}
\end{figure}

\textit{Disordered State} 
We start our analysis of the phase portrait around the disordered state $(\rho,\mathbf{p}_0) = (\rho_0,\mathbf{0})$. As expected from the stability analysis, the disordered state is stable for $\zeta<\zeta^c$ and $\Gamma_0>\Gamma_0^c$, which encompasses region $D$ of the phase diagram figure \ref{fig:phase_portrait}. We do not include snapshots of the dynamics in figure \ref{fig:snapshots}, since there are no notable dynamics or features to report.\\

\textit{Isotropic Turbulence}
For $\zeta\ll\zeta^c$ and $\Gamma_0<\Gamma_0^c$ Isotropic Turbulence is observed (region IT in figure \ref{fig:phase_portrait}). This state is governed by vortex-like structures which split and merge but exhibit a characteristic length scale, see figure \ref{fig:snapshots}\textbf{A}. This length scale can be obtained by the dominant mode of Eq.\ \eqref{eq:sigma1}, which reveals $\Lambda=2\pi \sqrt{-2\Gamma_2/\Gamma_0}$. Numerically, the length scale manifests itself as a dip in the spatial (time averaged) velocity correlation function and as a peak in the power spectrum (not plotted here, see \cite{Wensink2012a,Dunkel2013b}) as these quantities are linked by the Wiener-Khinchin theorem \cite{frisch1995turbulence}. The density stays almost constant throughout the simulation (narrow distribution around $\rho_0$, see figure \ref{fig:snapshots}\textbf{A} and Appendix \ref{app:phase_identifiers}). As briefly discussed in the introduction of this section, we label this state as Isotropic Turbulence (IT) to account for the absence of jets commonly found in bacterial turbulence, see \cite{Dunkel2013b}.\\

\textit{Motility-Induced Clustering and Motility-Induced Phase Separation}
Taking $\zeta\geq\zeta^c$ and $\Gamma_0\gg\Gamma_0^c$ leads to Motility-Induced Clustering (MIC) and Motility-Induced Phase Separation (MIPS), to be found in region MIC of the phase portrait. The dynamics are characterized by the emergence of dense clusters with $\mathbf{v}=0$ surrounded by a dilute phase, see figure \ref{fig:snapshots}\textbf{B}. We term the generic case MIC, but refer to MIPS when cluster coarsen over time, eventually reaching a completely phase-separated state, see figure \ref{fig:snapshots}\textbf{D}. In MIPS, clusters have an almost perfect spherical shape and their number decreases monotonically.\\

\textit{Isotropic Turbulence with Clustering}
In the lower right corner of the phase portrait (region ITC), a combination of the two states discussed previously is encountered. To be precise, we observe a phase separation into a dense phase with $\mathbf{v}=0$ and a dilute phase with $\mathbf{v}\neq0$. The dilute phase shows dynamics similar to isotropic turbulence. That is, we encounter vortices with a characteristic length scale in the dilute part of the simulation domain. Snapshots can be seen in \ref{fig:snapshots}\textbf{C} and \textbf{E}. Note that the interfaces between dilute and dense phases are highly irregular. Furthermore, we observe fluctuations in the number of clusters and their shape.\\

Most of the dynamics discussed previously can be expected from the linear stability analysis. However, there are two noteworthy exceptions which we label as regions $\text{ITC}_2$ and $\text{ITC}_3$ in figure \ref{fig:phase_portrait}. To understand the dynamics in these regimes, we have to study the nucleation and coarsening processes in more detail.

\subsubsection{Nucleation through turbulence}
\begin{figure}[!h]
	\centering
	\includegraphics[width=\linewidth, height=\textheight,keepaspectratio]{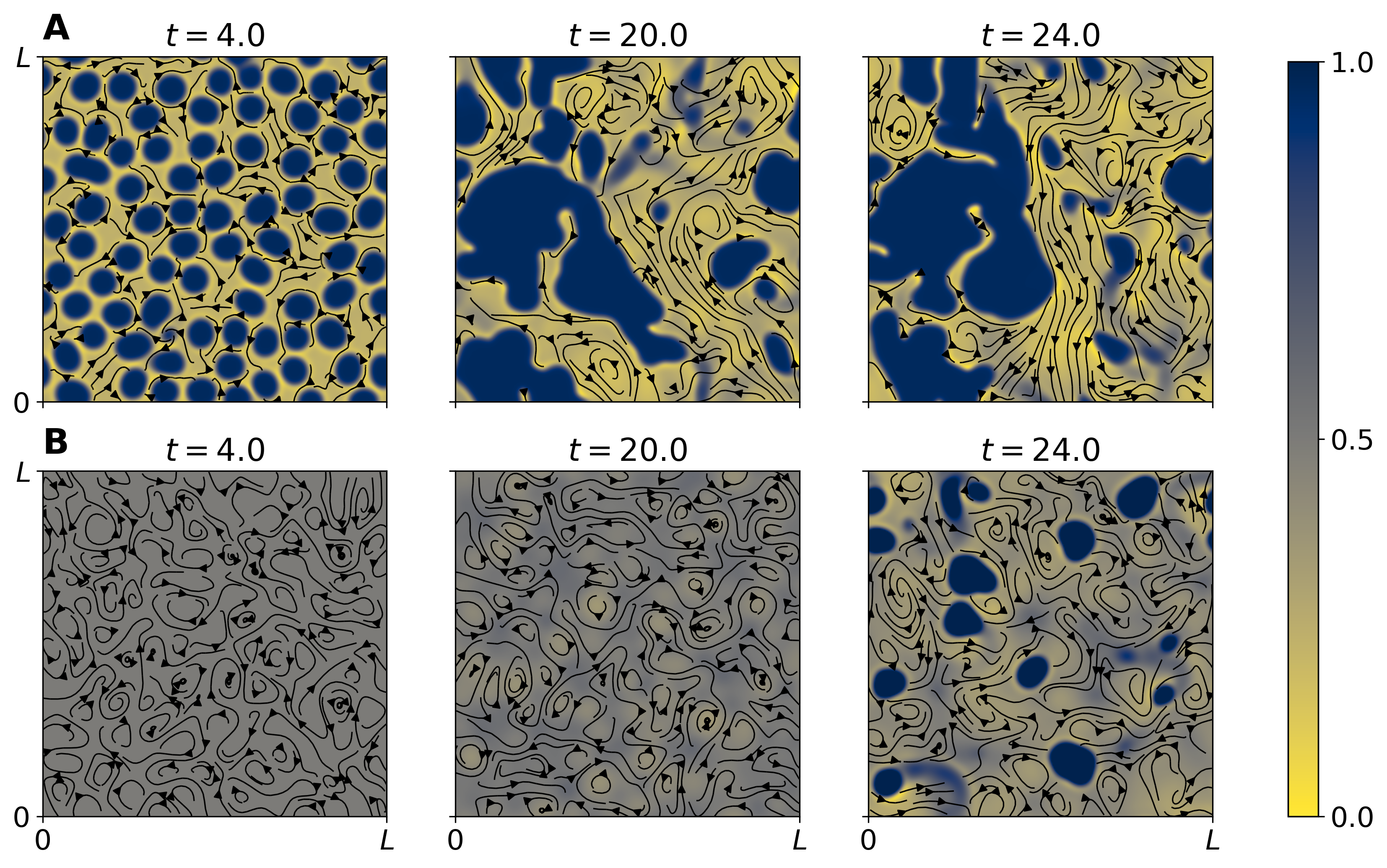}
	\caption{Nucleation dynamics \textbf{A} Spinodal decomposition precedes turbulence (region $\text{ITC}_1$) \textbf{B} Nucleation (and growth) is triggered by density inhomogenities due to the turbulent motion (region $\text{ITC}_2$). Depicted are the same simulation runs as in figure \ref{fig:snapshots} \textbf{D} and \textbf{C} respectively at $t=4,20,24$.}
	\label{fig:snapshots_nucleation}
\end{figure}

Insights into region $\text{ITC}_2$ can be gained by studying the nucleation of clusters. Nucleation and coarsening in MIPS is well studied in the literature, see for example \cite{Speck2014,Speck2015,Gonnella2015,Stenhammar2014,Wittkowski2014,Stenhammar2013,Patch2017}. A central result is that these processes in MIPS are quite similar to passive gas-liquid phase separation. As this classical transition is known to be a first-order phase transition (with the critical point being a notable exception) the same applies for MIPS, see \cite{Levis2017,Solon2018a} for an extensive study. Accordingly, clustering and phase separation can occur via two different mechanisms: Either nucleation and growth (in the metastable region) or spinodal decomposition (in the spinodal region). In the latter case there is no energy barrier to form a new phase. Hence, small perturbations start growing almost instantly. Therefore, the boundaries of the spinodal region, also known as the spinodal, can be detected by a linear stability analysis. In our case this corresponds to vertical yellow line in figure \ref{fig:phase_portrait}. To the right of that line we observe spinodal decomposition, either with or without the presence of turbulence, see figure \ref{fig:snapshots_nucleation}\textbf{A}.

However, the spinodal region is accompanied by a metastable region. There, perturbations have to overcome an energy barrier to form nuclei. In classic gas-liquid phase separation this energy barrier can be overcome over time by thermal fluctuations. As this is not possible in our model, we do not observe nucleation and growth in the absence of turbulence. In contrast, the turbulent motion for $\Gamma_0<\Gamma_0^c$ leads to local density inhomogeneities, which can overcome the energy barrier and trigger nucleation. This is what is found in region $\text{ITC}_2$ and illustrated in figure \ref{fig:snapshots_nucleation}\textbf{B}. At $t=4$, no clustering is observed, whereas for spinodal decomposition clusters appear throughout the entire simulation domain, see figure \ref{fig:snapshots_nucleation}\textbf{A}. Turbulence sets in at approximately $t\approx20$, which leads to the formation of a few nuclei at random sites, see $t=24$. We want to point out that this is a surprising result. Since inertial turbulence is associate with mixing, one would expect turbulence to inhibit nucleation.

This mechanism offers a possible explanation for the shape of the phase boundary between regions IT and $\text{ITC}_2$. Clearly, nucleation and growth are only possible in the metastable regime. The extent of the metastable region can be numerically estimated by checking for hysteresis. We initiate simulation runs either with a homogeneous density profile or with a completely phase-separated state consisting of a single droplet (in the non-turbulent regime). The lower boundary of the metastable region is reached when the droplet looses stability. From our simulations this point can be estimated to be at $\zeta_b \approx 0.68$. The upper boundary is provided by the spinodal at $\zeta^c=1.25$. While checking for hysteresis, we made another important observation. Close to the spinodal a droplet with slightly higher density than the average density $\rho_0$ is sufficient to trigger nucleation. At the binodal, only droplets with a density close enough to the maximal density are stable. Additionally, the maximal density increases when decreasing $\zeta$ due to Eq.\ \eqref{eq:v_rho_linear}. Hence, when moving away from the spinodal, larger density inhomogeneities have to be provided to trigger nucleation. Furthermore, from our simulations we observe that the amplitude of density variations, i.e.\ $\rho_{max}-\rho_{min}$, is proportional to $\Gamma_0^c-\Gamma_0$, see Appendix \ref{app:di}. Hence, for $\Gamma_0\approx\Gamma_0^c$, only small density inhomogenities are observed. These are enough to trigger nucleation close to the spinodal, but are not sufficient close to the binodal. Altogether, these observations explain the shape of the phase boundary in the metastable regime of the phase portrait.

\subsubsection{Altered Ostwald Ripening}
\begin{figure}[!h]
	\centering
	\includegraphics[width=\linewidth, height=\textheight,keepaspectratio]{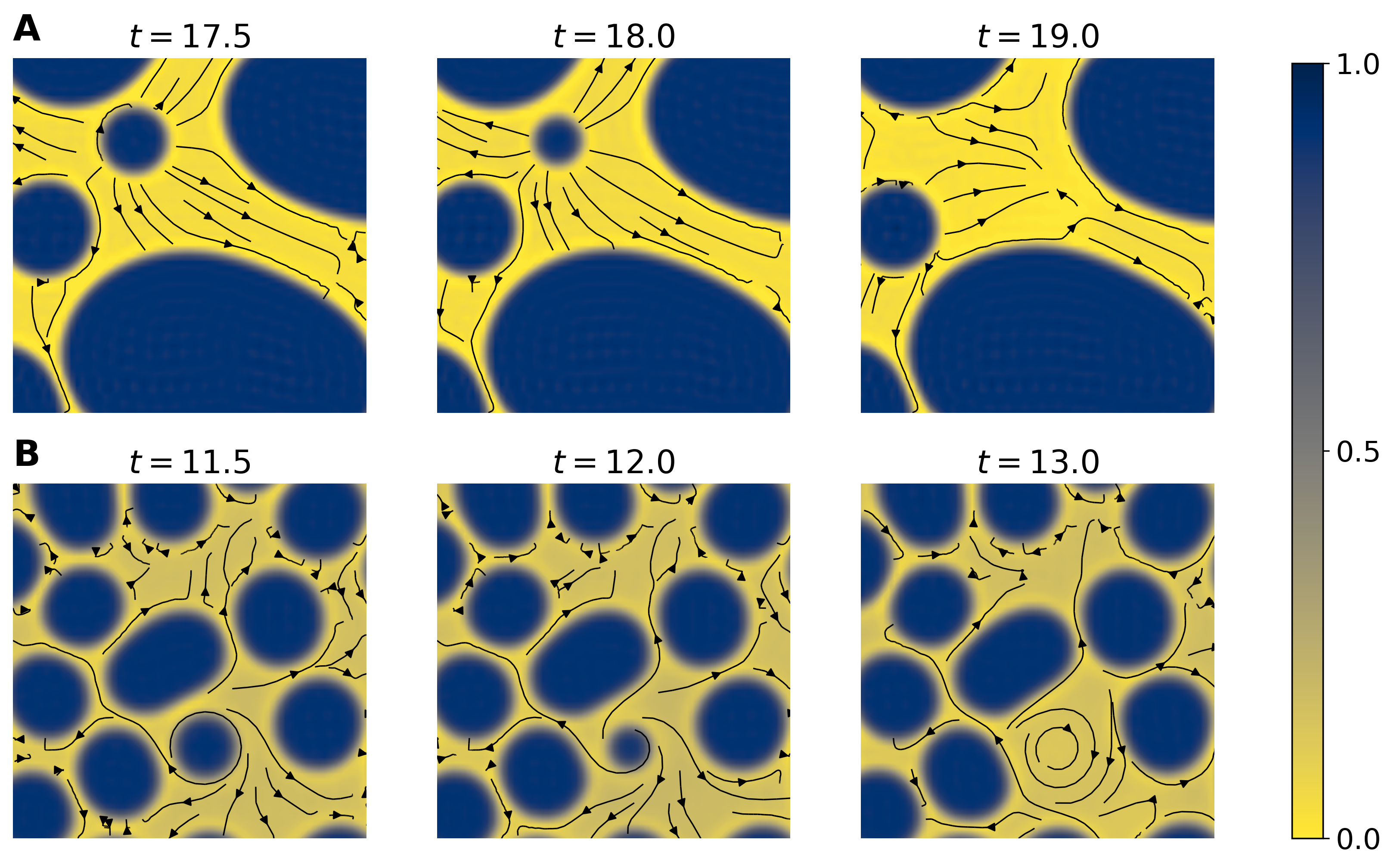}
	\caption{Ostwald Ripening without (\textbf{A}) and with (\textbf{B}) the convective term $\lambda_0(\mathbf{p}\cdot\nabla)\mathbf{p}$. Streamlines are perpendicular to the surface of the dissolving cluster for \textbf{A} whereas they are tangential in \textbf{B}. The remaining parameters are chosen as listed in table \ref{tab:sim_parameters_interface} in Appendix \ref{app:sim_parameters}.}
	\label{fig:snapshots_interface}
\end{figure}

We will now focus on region $\text{ITC}_3$ of figure \ref{fig:phase_portrait}, i.e.\ we want to answer the question why we observe enstrophy above the critical value for the onset of turbulence $\Gamma_0^c$. To explain this, we focus on the coarsening kinetics. Coarsening in MIPS is similar to Ostwald ripening, see \cite{Cates2010,Gonnella2015,Stenhammar2014}. That is, clusters grow on the expense of smaller clusters, which dissolve and redeposit onto larger ones. What that process typically looks like for classical MIPS (model \eqref{eq:MIPS}) is shown in figure \ref{fig:snapshots_interface}\textbf{A}. During the dissolution process, the mass flux streamlines (which indicate the direction of mass transport) are perpendicular to the dissolving surface. Almost immediately after the cluster disappears, the streamlines rearrange, leaving no trace of the dissolved cluster, see $t=19$.

However, the presence of the convective term $\lambda_0(\mathbf{p}\cdot\nabla)\mathbf{p}$ in our simulations (see model \eqref{eq:Dynamics}) seems to alter the dissolution dynamics. The snapshots in figure \ref{fig:snapshots_interface}\textbf{A} are produced by setting $\lambda_0=0$, whereas the snapshots in figure \ref{fig:snapshots_interface}\textbf{B} are obtained for $\lambda_0=3$ (all other parameters are unchanged). For the latter choice, the streamlines appear tangential to the cluster surface, leading to a vortex after the cluster disappears. As $\lambda_0=3$ is the parameter used to compute the phase diagram, this behaviour is observed in the entire region ITC. In region $\text{ITC}_3$ (as opposed to regions $\text{ITC}_1$ and $\text{ITC}_2$), these vortices are transient, i.e.\ they vanish some time $\Delta t$ after the cluster has dissolved. This is expected as the finite-wavelength instability is inactive in this regime, i.e.\ turbulence is not self-sustained. However, as other clusters dissolve, new vortices are formed constantly, leading to an increased enstrophy over a long time, see figure \ref{fig:add_figs} in Appendix \ref{app:di}.

While focusing on coarsening dynamics, we made another interesting observation concerning the phase portrait: For some simulation runs, there appears to be no coarsening (at least on the simulation time scale). That is, the number of clusters does not decrease nor does the size of the largest cluster increase, see Appendix \ref{app:coarsening} for details. The dynamics for these cases are shown in figure \ref{fig:snapshots}\textbf{B} and \textbf{C} without and with the presence of turbulence respectively.

Altogether, the linear stability analysis provides a viable intuition into the expected phases. However, it fails to cover metastable regimes and nonlinear effects. Nevertheless, these effects and the apparent arrest of coarsening pose an intriguing research opportunity, which we will pursue in the future.

\subsection{Anomalous velocity statistics}\label{sec:anomalous}

\begin{figure}[!h]
    \centering
    \includegraphics[scale=0.5]{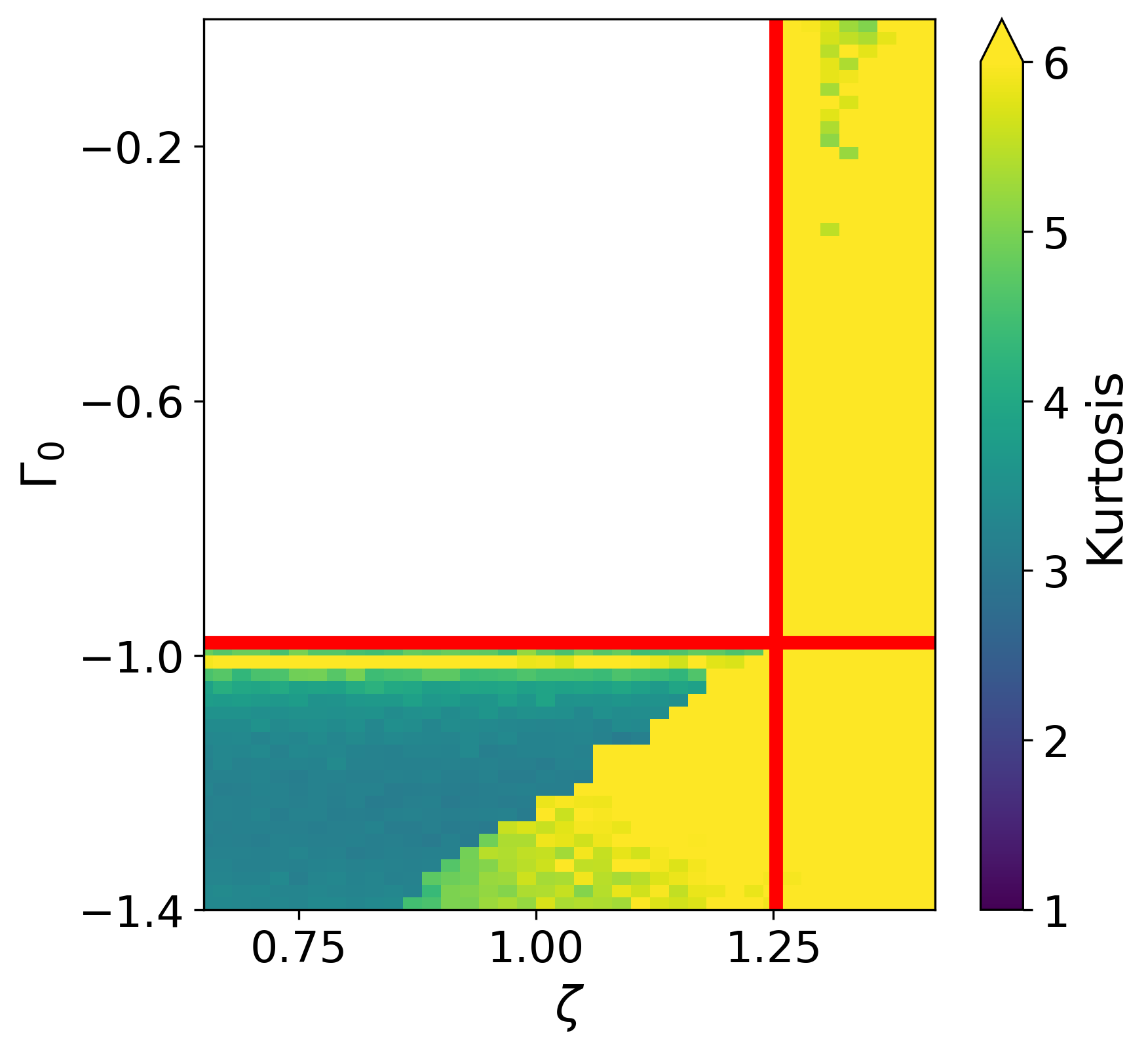}
    \caption{Kurtosis in the $\zeta$-$\Gamma_0$ plane. For $\zeta<\zeta_c$ and $\Gamma_0>\Gamma_0^c$ no velocity statistics are computed as $\mathbf{v}\equiv0$ (up to numerical errors). Simulation runs with a kurtosis higher than 6 are yellow.}
	\label{fig:phase_portrait_kurtosis}	
\end{figure}

The main motivation to study the combined model \eqref{eq:Dynamics} was to observe and explain anomalous velocity statistics. To quantify the deviations from Gaussian statistics, we compute the kurtosis $\kappa$ of the standardized velocities $\hat v_i = (v_i-\overline{\langle v\rangle})/\sigma(v_i)$, where $\overline{\langle v\rangle}$ and $\sigma(v_i)$ are the mean and the standard deviation of the velocity component $v_i$ respectively. The kurtosis coincides with the fourth moment for a standardized distribution. In the lower left corner of figure \ref{fig:phase_portrait_kurtosis} (almost) normal statistics are observed, i.e. a kurtosis $\kappa\approx3$ is reported. This is not surprising since this region coincides with the turbulent regime (region IT in figure \ref{fig:phase_portrait}), where we expect normal statistics as in the incompressible model \eqref{eq:Incomp}. However, strongly anomalous statistics ($\kappa\geq6$) are reported for a large part of the phase portrait. The anomalous region matches with the clustering phases (ITC and MIC) of figure \ref{fig:phase_portrait}, indicating that the deviation from normal statistics stems from clustering.

\begin{figure}[!h]
	\centering
	\includegraphics[width=\linewidth, height=\textheight,keepaspectratio]{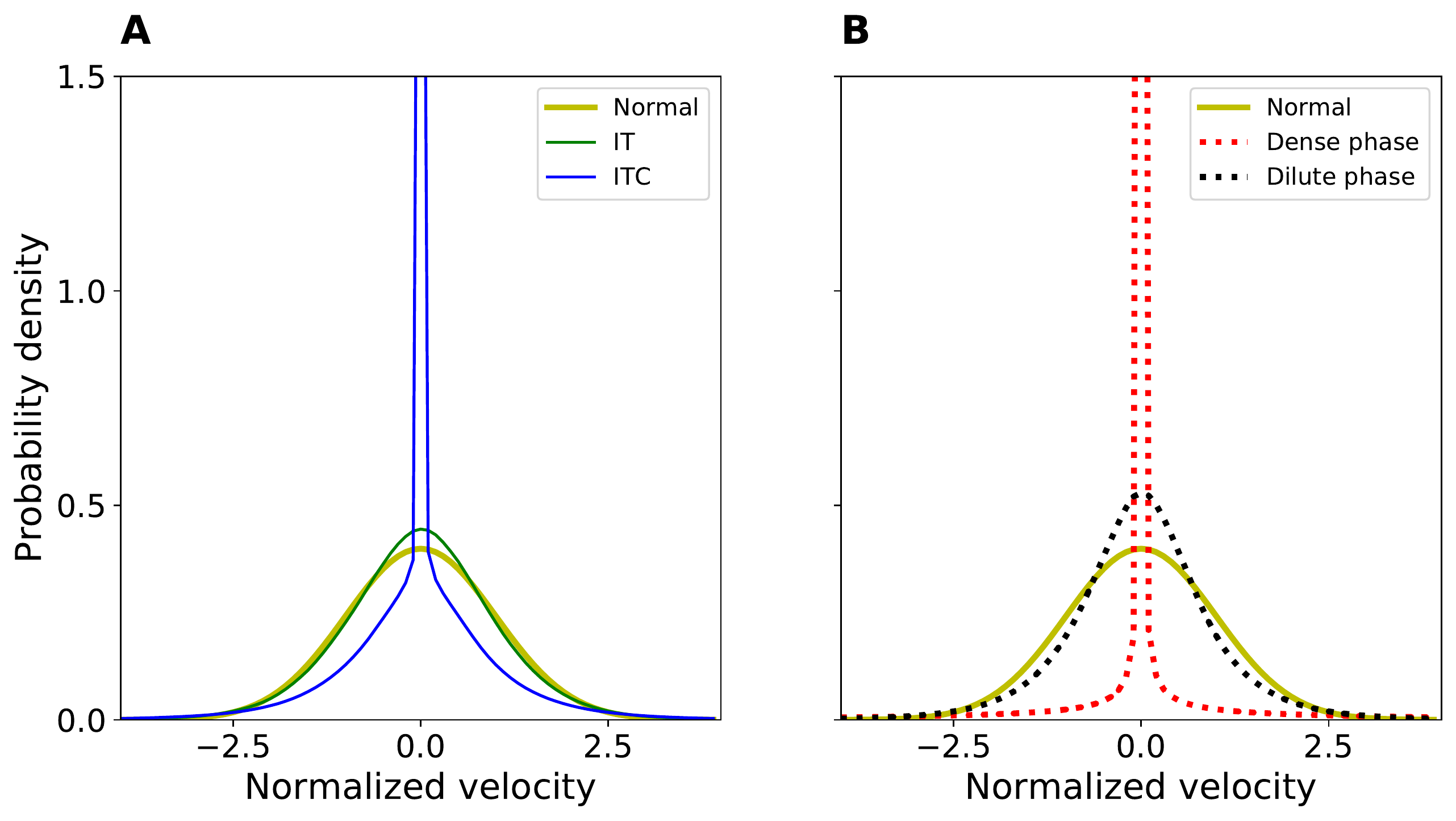}
	\caption{Velocity statistics \textbf{A} Typical distribution functions of the standardized velocity components for Isotropic Turbulence (IT) and Isotropic Turbulence with Clustering (ITC) and a normal distribution as comparison. \textbf{B} Velocity distribution for the dense and dilute regions (dotted lines) for Isotropic Turbulence with Clustering. Data obtained as described in Appendix \ref{app:sim_parameters}.}
	\label{fig:anomalous}
\end{figure}

Indeed, that hypothesis is supported by investigating the velocity statistics in more detail. Compared to the (almost) normal statistics in the IT regime, a clear peak around zero is visible in the ITC phase in figure \ref{fig:anomalous}\textbf{A}. This peak can be explained by a subpopulation argument: Computing the velocity statistics in the dense clusters and dilute, turbulent regimes separately reveals a clear split, see figure \ref{fig:anomalous}\textbf{B}. Details on the thresholding can be found in Appendix \ref{app:coarsening}. As expected, the statistics in the dilute part are similar to the ones reported for IT, whereas in the dense phase $\mathbf{v}\approx 0$. Combining both distributions gives the blue curve in figure \ref{fig:anomalous}\textbf{A}. Note that the statistics in the dilute phase are not perfectly Gaussian. The offset could be due to the interface between phases and the interactions of turbulence and clustering. Moreover, the statistics in the IT phase already show a slight deviation from normal statistics, i.e.\ they are only approximately Gaussian.

Furthermore, we want to point out that figure \ref{fig:anomalous}\textbf{B} was produced in the spinodal regime (region $\text{ITC}_1$), i.e.\ when clusters form due to spinodal decomposition. The subpopulation argument works worse in the metastable regime, i.e.\ when nucleation is triggered by turbulence (region $\text{ITC}_2$). We speculate that in this case the interactions between clustering and turbulence are pronounced, leading to stronger correlations. Also note that the transition from normal to anomalous statistics gets smeared out for lower values of $\zeta$. This hints at a possible continuous transition at the nucleation point. Altogether, the statistical properties of the system in the metastable regime require a more fundamental analysis and deeper understanding of the interactions in the competing processes.

\section{Conclusion, Discussion and Outlook}\label{sec:discussion}

We have presented and studied a phenomenological model combining ideas from MIPS and a fourth-order theory proposed to describe active turbulence. The underlying theories are distinct in the type of main instability they describe. While MIPS is driven by a global (long-wavelength) instability, meso-scale turbulence is characterized by a specific length scale (short-wavelength instability). Our model inherits both instabilities, which results in rich dynamics and an interesting interplay between the two.

While we showed that turbulence can trigger nucleation, the effect of turbulence on the coarsening process is not yet clear. Important alterations concerning the shape of the phase boundaries, fluctuations and cluster fluidity are evident from our simulations. However, these aspects need to be investigated further to quantify their importance and physical origin. Furthermore, we observed situations where coarsening appears to be frozen on a certain length scale. As this happens also in the absence of turbulence, we speculate that such an arrested phase separation might be connected to the loss of pure relaxation dynamics by the convective term.


Spatial segregation into static clusters surrounded by swarming bacteria was reported for colonies of \textit{Bacillus subtilis} when adding (sublethal concentrations of) antibiotics in \cite{Benisty2015}. Motile cells are diluted, resulting in a lower density of swarming bacteria. Our simulations show qualitatively similar results as clusters with $\mathbf{v}=0$ are surrounded by a dilute, turbulent phase. Moreover, we observe anomalous statistics with a kurtosis up to 6 or larger, which was also reported in \cite{Benisty2015}. Furthermore, a recent experimental study \cite{Grobas2020} indicates that the coexistence of swarming dynamics and MIPS might explain the stress-induced transition to biofilm.

Connecting phenomenological transport coefficients with experimentally traceable parameters is not straightforward. The derivation of model \eqref{eq:Incomp} presented in \cite{Heidenreich2016,Reinken2018a} can give a hint for most parameters. However, obtaining the exact form of $v(\rho)$ for biological system from microscopic considerations is fairly complicated. Alternatively, the local $v(\rho)$ can be fitted to (global) experimental data reported in \cite{Ariel2018,Aranson2007} for example. This data suggests a non-monotone dependence of velocity on density, in contrast to the simple linear assumption we used in our numerical study. Nevertheless, preliminary simulations show that our general results still hold for more complicated functions $v(\rho)$.

Our phenomenological approach allows us to understand anomalous statistics in meso-scale turbulent systems without specifying the microscopic details. Hence, our model might be applicable for different experimental setups. For example, anomalous velocity statistics were reported in \cite{Ilkanaiv2017} (different aspect ratio of mutated \textit{Bacillus subtilis}) and \cite{Beer2020} (monolayer swarming of \textit{Bacillus subtilis} for low density). However, different mechanisms could be underlying these observations. For example, a chemical response could lead to velocity variations while leaving the density unchanged. Such a situation could be possibly modelled by allowing $v(\rho)$ to depend on an external scalar field instead of the density.

Altogether, our results provide a simple explanation for how anomalous statistics can arise in meso-scale turbulence. While addressing that topic, new questions are raised concerning the role of turbulence for nucleation and coarsening. This poses a challenging research opportunity we will pursue in the future.

\section*{Acknowledgments}
We are grateful to Henning Reinken, Michael Wilczek and Nir Gov for helpful discussion. This work was supported by the Deutsche Forschungsgemeinschaft (DFG) through grants HE 5995/3-1 (SH, VMW and AB), BA 1222/7-1 (MB and GA) and SFB 910 (projects B4 (HS) and B5 (MB)). GA and AB are thankful for partial support from the Israel Science Foundation grant 373/16.

\clearpage
\appendix

\section{Stability Analysis}\label{app:stability}
The stability analysis of the system \eqref{eq:Dynamics} can be done in quite general fashion, i.e. without specifying the exact from of the coupling terms $v(\rho)$ and $A(\rho)$ nor the steady state. Introducing perturbations $\mathbf{p}\mapsto\mathbf{p}_0 + \delta\mathbf{p}$ and $\rho \mapsto \rho_0 + \delta\rho$ and linearizing Eq.\ \eqref{eq:Dynamics} close to the steady state $(\rho,\mathbf{p}) \equiv (\rho_0,\mathbf{p}_0)$ yields the equations of motion for the perturbations
\begin{equation}
	\begin{aligned}
		\partial_t \delta\rho = &-v'(\rho_0)\mathbf{p}_0\cdot\nabla\delta\rho + D\Delta\delta\rho -v(\rho_0)\nabla\cdot\delta\mathbf{p},\\
		\partial_t \delta\mathbf{p} = &-\frac{1}{2}\left(v(\rho_0) + v'(\rho_0)\rho_0\right) \nabla\delta\rho - A'(\rho_0)\mathbf{p}_0\delta\rho
		-\lambda_0((\mathbf{p}_0\cdot\nabla)\delta\mathbf{p})\\
		&-A(\rho_0)\delta\mathbf{p} - C\left(2\mathbf{p}_0\mathbf{p}_0 + |\mathbf{p}_0|^2 I\right)\delta\mathbf{p} + \Gamma_0\Delta\delta\mathbf{p} -\Gamma_2\Delta^2\delta\mathbf{p}.
	\end{aligned}
\end{equation}
Expanding the perturbations in Fourier modes with $\mathbf{k}=(k_1,k_2)$, i.e. setting
\begin{equation}
    \begin{pmatrix}
        \delta\rho\\\delta\mathbf{p}
    \end{pmatrix}
    =\begin{pmatrix}
        \hat\rho\\\mathbf{\hat p}
    \end{pmatrix}
    \exp{(\sigma t+i\mathbf{k}\cdot\mathbf{x})}
\end{equation}
leads to the coupled dispersion relations
\begin{equation}\label{eq:Dispersion_General}
    \begin{aligned}
    \sigma\hat\rho = &-(i v'(\rho_0)\mathbf{p}_0\cdot\mathbf{k} + D|\mathbf{k}|^2)\hat\rho -i v(\rho_0)\mathbf{k}\cdot\mathbf{\hat p},\\
    \sigma\mathbf{\hat p} = &-\left(i\frac{1}{2}\left(v(\rho_0) + v'(\rho_0)\rho_0\right) \mathbf{k} + A'(\rho_0)\mathbf{p}_0\right)\hat\rho\\
    &-\left(i\lambda_0(\mathbf{p}_0\cdot\mathbf{k}) +A(\rho_0) + C|\mathbf{p}_0|^2 + \Gamma_0|\mathbf{k}|^2 +\Gamma_2|\mathbf{k}|^4\right)\mathbf{\hat p}\\
    &- 2C\mathbf{p}_0\mathbf{p}_0\cdot\mathbf{\hat p},
    \end{aligned}
\end{equation}
where we grouped terms belonging to $\hat{\rho}$ and $\mathbf{\hat p}$. Hence, it is straightforward to read off the stability matrix from Eq.\ \eqref{eq:Dispersion_General}. Roughly speaking, a finite-wavelength instability is expected for sufficiently negative values of $\Gamma_0$ and a long-wavelength instability for $A(\rho_0)<0$ (disordered case). Another long-wavelength instability can be found in a similar way as in section \ref{sec:BaseModels} for MIPS: Inserting the first coupling term from the equation for $\mathbf{\hat p}$ into the equation for $\hat\rho$ results in an effective diffusion equation for the density similar to Eq.\ \eqref{eq:MIPS_effective}. For $v'(\rho_0)<0$ sufficiently small, the sign of the diffusion coefficient changes, triggering a long-wavelength instability.

As mentioned in the main text, solving for the eigenvalues of the stability matrix is possible, but yields complicated expressions which are not instructive. However, assuming a disordered state reduces the complexity of the calculation significantly. Rewriting Eq.\ \eqref{eq:Dispersion_General} in matrix notation for the disordered state $\mathbf{p}_0=0$ leads to the equation
\begin{equation}
	\sigma
	\begin{pmatrix}
	\hat\rho\\\mathbf{\hat p}
	\end{pmatrix}
	= \mathbf{M}(\mathbf{k})
	\begin{pmatrix}
	\hat\rho\\\mathbf{\hat p}
	\end{pmatrix}
\end{equation}
with stability matrix
\begin{equation}\label{eq:stability_matrix}
	\mathbf{M}(\mathbf{k})=
	\begin{pmatrix}
		&-D|\mathbf{k}|^2 &i\alpha k_1 &i\alpha k_2\\
		&i\beta k_1 &-A(\rho_0)-\Gamma_0|\mathbf{k}|^2-\Gamma_2|\mathbf{k}|^4 &0 \\
		&i\beta k_2 &0 &-A(\rho_0)-\Gamma_0|\mathbf{k}|^2-\Gamma_2|\mathbf{k}|^4
	\end{pmatrix},
\end{equation}
with coupling terms $\alpha, \beta$ given by
\begin{equation}\label{eq:coupling_stability}
	\alpha = -v(\rho_0), \qquad \beta=-\frac{1}{2}\left(v(\rho_0)+v'(\rho_0)\rho_0\right).
\end{equation}
Note that $\alpha\beta=\gamma$ for $\gamma$ introduced in Eq.\ \eqref{eq:coupling_combined}. The stability matrix Eq.\ \eqref{eq:stability_matrix} contains a scalar multiple of the identity as a block matrix in the lower right corner. This makes the computation of the eigenvalues much handier as compared to the polar case and leads to the dispersion relations Eq.\ \eqref{eq:Dispersion}.

\section{Phase Identifiers}\label{app:phase_identifiers}
The phase diagram figure \ref{fig:phase_portrait} is produced by solving Eq.\ \eqref{eq:Dynamics} and assigning a phase based on two quantifiers: the enstrophy of the polarization density and the modality of the density distribution. Runs are initiated with small perturbations around the overall density $\rho_0$ and a perturbed smooth polarization field with low amplitude.

\begin{figure}[!htb]
	\centering
	\begin{tabular}{c | c c}
		State & $\Omega$ & $\rho$ (pdf) \\ \hline
		Disordered & $\approx0$  & unimodal \\
		Isotropic Turbulence & $\gg1$ & unimodal\\
		Motility-Induced Clustering & $\approx0$ & multimodal\\
		Isotropic Turbulence with Clustering & $\gg1$ & multimodal\\
	\end{tabular}
	\caption{Identification of states according to identifiers: Rescaled enstrophy $\Omega$ (first row) and modalitiy of the probability density function of the (physical) density $\rho$.}
	\label{tab:phase_char}
\end{figure}

\begin{figure}[!htb]
	\centering
	\includegraphics[scale=0.55]{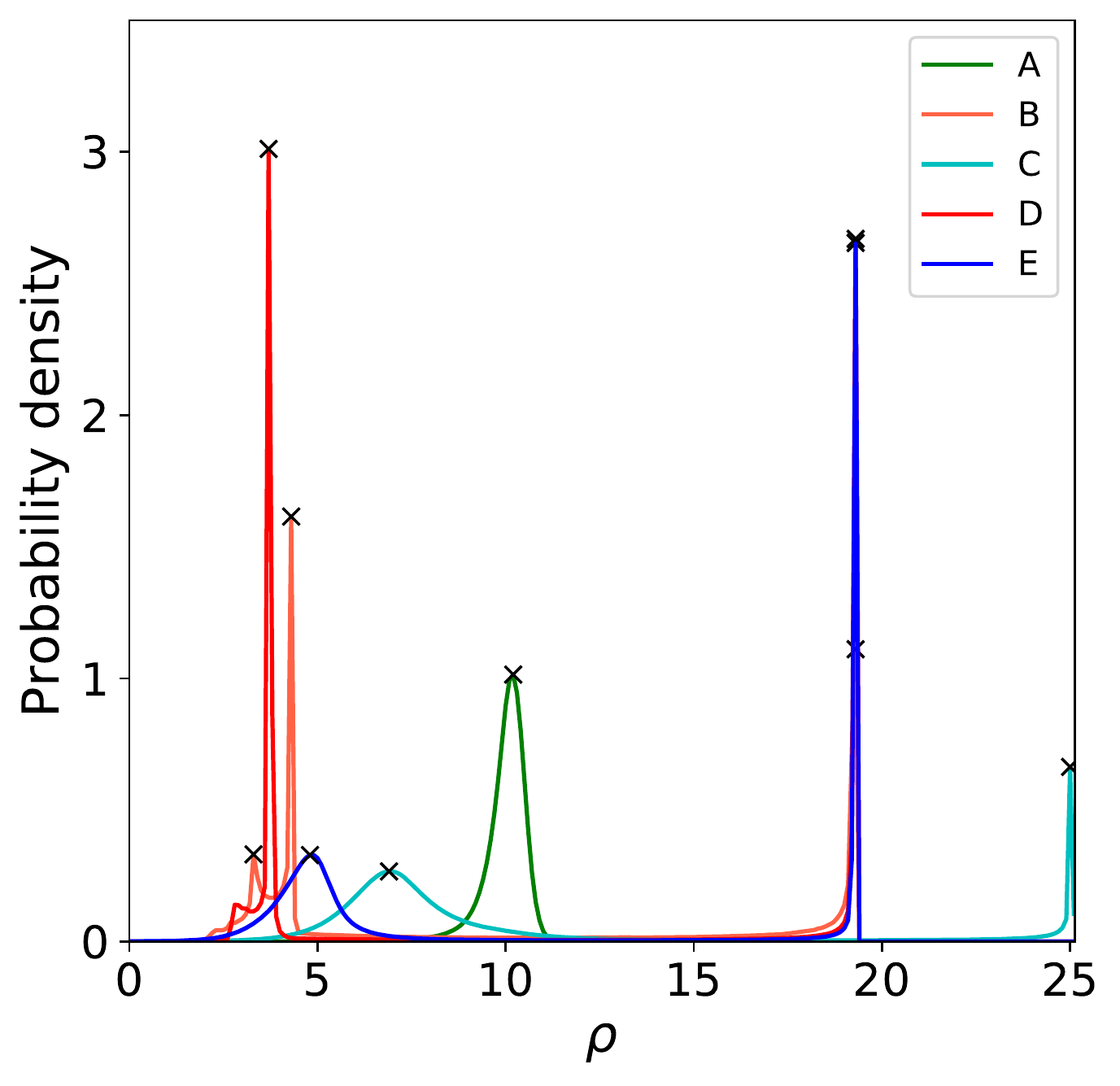}
	\caption{Non rescaled density distribution for the five cases with $\rho_0=10$ depicted in figure \ref{fig:snapshots}. Labels are in line with the subplots of figure \ref{fig:snapshots}. Detected density peaks are marked with a cross.}
	\label{fig:density_dist}
\end{figure}

The enstrophy is defined as $\Omega = \frac{1}{2}\overline{\langle|\omega(\mathbf{x},t)|^2\rangle}$, where $\omega$ is the vorticity field $\omega = \partial_x v_y - \partial_y v_x$ of some vector field $\mathbf{v}=(v_x,v_y)$. Brackets and overbars denote spatial and temporal averages respectively. We compute the enstrophy of the polarization density $\mathbf{p}$ in the dilute phase as a measure for the presence of vortices. The density distribution is computed after an initial transient and checked for bimodality. The transient is needed to filter out the initial (almost homogeneous) density field. Results are shown in figure \ref{fig:density_dist} for the five cases depicted in figure \ref{fig:snapshots}.

\section{Numerical Implementation}\label{app:numerics}
We solve the system \eqref{eq:Dynamics} numerically using a pseudo-spectral method with operator splitting for time integration \cite{Kozlov2008}. Nonlinearities are treated by applying a 2/3 dealias rule \cite{canuto2007spectral}. While pseudo-spectral methods have been proven to be powerful and reliable for incompressible flows, special care is needed when dealing with conservation laws \cite{Ghosh1993}. The main culprit lies in sharp boundaries between phases. Those are ill-suited for a global method and result in numerical errors in the form of Gibbs phenomena \cite{Gottlieb1997,Gelb1997}. This necessitates very fine grids. As we are studying coarsening kinetics (long time scale, large system size) finite computation resources and times restrict the resolution of the grid. A more feasible and numerically efficient approach is choosing the diffusion coefficient of the density equation big enough to sufficiently smooth out phase boundaries. This allows us to choose grid sizes ranging from 128x128 to 512x512 and time steps of $\Delta t = 10^{-3}\dots 10^{-1}$. Alternatively, advanced techniques \cite{Abarbanel1985,Gottlieb2005} could be applied to increase stability and accuracy.

\section{Simulation Parameters}\label{app:sim_parameters}
Figure \ref{fig:Disp} is created using the parameters listed in table \ref{tab:dispersion_parameters}. For \textbf{A}-\textbf{C} we set $A=0.2$ (disordered state) and for \textbf{D}-\textbf{F} we choose $A=-0.1$ (polar state). Additionally, for \textbf{A} and \textbf{D} we set $\Gamma_0=-1.2,\zeta=1.95$, for \textbf{B} and \textbf{E} we set $\Gamma_0=-0.4,\zeta=2.01$, for \textbf{C} and \textbf{F} we set $\Gamma_0=-1.2,\zeta=2.01$.

\begin{table}[!h]
    \centering
    \begin{tabular}{c c c c c c}
        $v_0$ & $\rho_0$ & $D$ & $\lambda_0$ & $C$ & $\Gamma_2$\\ \hline
        20 & 5 & 1 & 3 & 1 & 0.8
    \end{tabular}
    \caption{Simulation parameters for figure \ref{fig:Disp}.}
    \label{tab:dispersion_parameters}
\end{table}

Figures \ref{fig:phase_portrait}, \ref{fig:snapshots}, \ref{fig:phase_portrait_kurtosis}, \ref{fig:anomalous}, \ref{fig:density_dist} and \ref{fig:add_figs} are all produced with the parameter set given in table \ref{tab:sim_parameters}. Additional parameters are provided in the following.

\begin{table}[!h]
    \centering
    \begin{tabular}{c c c c c c c c}
        $v_0$ & $\rho_0$ & $D$ & $\lambda_0$ & $A$ & $C$ & $\Gamma_2$ & $L$\\ \hline
        25 & 10 & 1 & 3 & 0.3 & 0 & 0.8 & 16$\pi$
    \end{tabular}
    \caption{Simulation parameters for figures \ref{fig:phase_portrait}, \ref{fig:snapshots}, \ref{fig:phase_portrait_kurtosis}, \ref{fig:anomalous}, \ref{fig:density_dist} and \ref{fig:add_figs}.}
    \label{tab:sim_parameters}
\end{table}

The subplots in figure \ref{fig:snapshots} are generated according to table \ref{tab:sim_parameters_snapshots}. Figure \ref{fig:snapshots_nucleation} \textbf{A} and \textbf{B} show runs depicted in figure \ref{fig:snapshots} \textbf{D} and \textbf{C} at earlier times respectively. Furthermore, the data for figure \ref{fig:anomalous} is obtained from the simulation runs shown in figure \ref{fig:snapshots} \textbf{A} and \textbf{D} for IT and ITC.

\begin{table}[!h]
    \centering
    \begin{tabular}{c | c c c c c}
         & A & B & C & D & E \\ \hline
        $\zeta$ & 0 & 1.3 & 1.0 & 1.3 & 1.3 \\
        $\Gamma_0$ & -1.4 & 0.1 & -1.4 & 1.4 & -1.4  
    \end{tabular}
    \caption{Additional parameters for the subplots of figure \ref{fig:snapshots}.}
    \label{tab:sim_parameters_snapshots}
\end{table}

Simulation parameters for figure \ref{fig:snapshots_interface} are given in table \ref{tab:sim_parameters_interface}. For figure \ref{fig:snapshots_interface}\textbf{A} we set $\lambda_0=0$, whereas for \textbf{B} we choose $\lambda_0=3$.

\begin{table}[!h]
    \centering
    \begin{tabular}{c c c c c c c c c}
        $v_0$ & $\zeta$ & $\rho_0$ & $D$ & $A$ & $C$ & $\Gamma_0$ & $\Gamma_2$ & $L$\\ \hline
        20 & 2.2 & 5 & 1 & 0.2 & 0 & 0.1 & 1 & 16$\pi$
    \end{tabular}
    \caption{Simulation parameters for figure \ref{fig:snapshots_interface}.}
    \label{tab:sim_parameters_interface}
\end{table}

\section{Additional figures for region $\text{ITC}_2$ and $\text{ITC}_3$}\label{app:di}
\begin{figure}[!htb]
    \centering
    \includegraphics[width=\linewidth, height=\textheight,keepaspectratio]{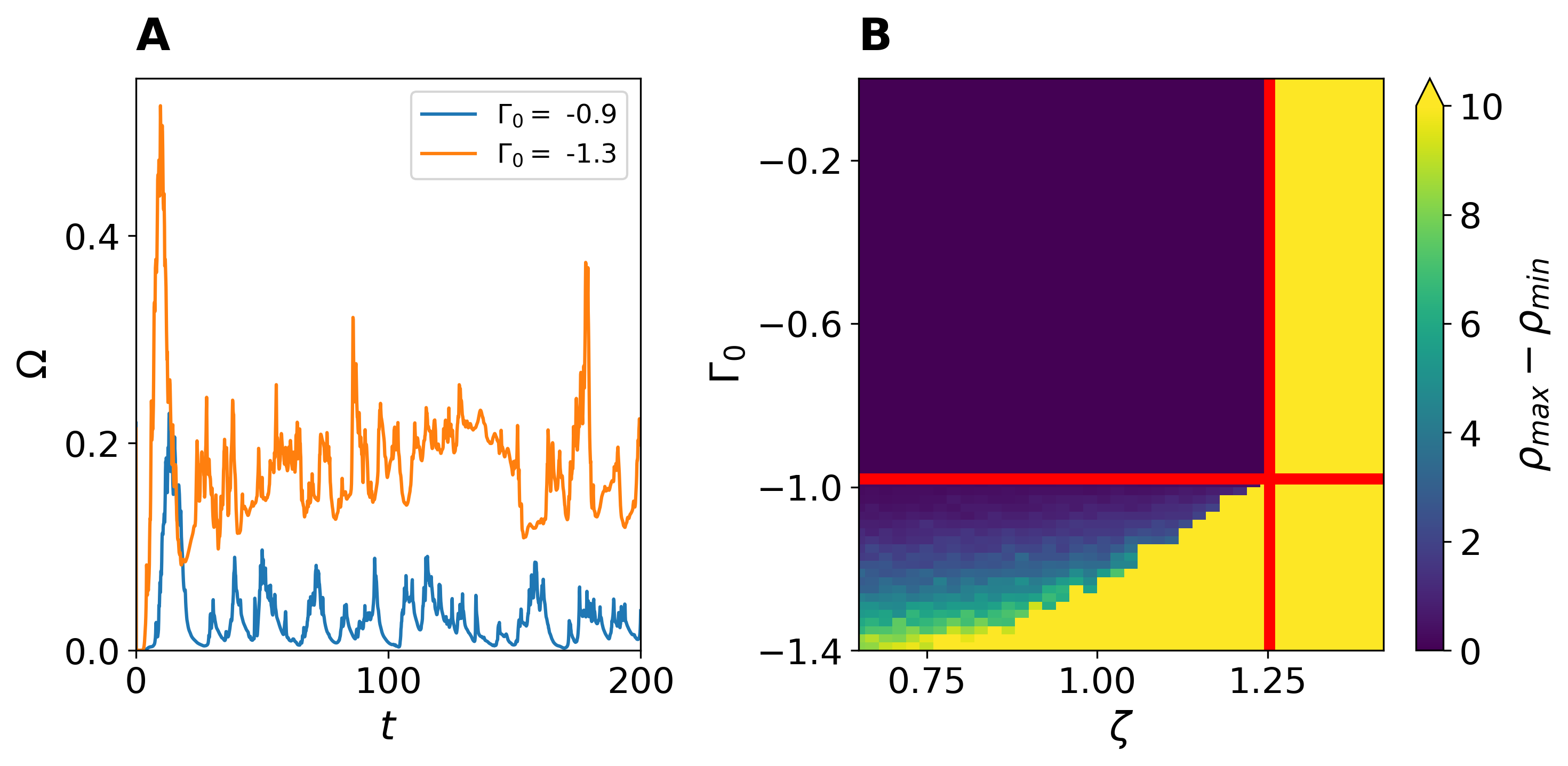}
    \caption{\textbf{A} Enstrophy $\Omega(t)$ over time for runs in region $\text{ITC}_1$ ($\Gamma_0=-1.3$) and region $\text{ITC}_3$ ($\Gamma_0=-0.9$). \textbf{B} Maximal density fluctuations $\rho_{max}-\rho_{min}$ as function of $\zeta$ and $\Gamma_0$.}
	\label{fig:add_figs}	
\end{figure}

In our simulations we observe the relationship $\rho_{max}-\rho_{min}\sim\Gamma_0^c-\Gamma_0$ between the amplitude of density variations and the distance to the critical parameter for the onset of turbulence $\Gamma_0^c$. This can be deduced from figure \ref{fig:add_figs}\textbf{B}. The values $\rho_{min}$ and $\rho_{max}$ are computed for every simulation run over the whole time and space domain, i.e.\ $\rho_{min} =\min\rho(\mathbf{x},t)$ and $\rho_{max} =\max\rho(\mathbf{x},t)$. Furthermore, we plot the time-dependent enstrophy $\Omega(t)$ for runs in region $\text{ITC}_1$ and $\text{ITC}_3$ in figure \ref{fig:add_figs}\textbf{A}. This shows that in region $\text{ITC}_3$ meso-scale turbulence is not self-sustained as $\Omega(t)\approx0$ for some $t$, while $\Omega(t)\gg0$ in region $\text{ITC}_1$.

\section{Coarsening}\label{app:coarsening}
We track the number of dense clusters and the size of the largest one to examine coarsening. The detection of a cluster is done as follows: We define a density threshold as
\begin{equation}
    \rho_{c} = \frac{\rho_0 + \rho_{max}}{2},
\end{equation}
where the maximal density is given by $\rho_{max} = v_0/\zeta$ by Eq.\ \eqref{eq:v_rho_linear}. This threshold is sufficient if the interfaces are negligible. After each snapshot is labeled according to this threshold, we use methods from image recognition \cite{scikit-image} to count and characterize clusters where special care is taken to account for the periodic boundary conditions. From this data linear trends are fitted for the number and mass fraction (mass of the cluster compared to the overall mass in the computation box). Since we are interested in the long time dynamics and not the onset of clustering, we start to fit after a transient of five time units since the emergence of the first cluster. We define a relatively generous criterion for coarsening by requiring non increasing numbers of clusters and an increase in mass of the largest cluster (applied to the linear fits).

\normalem
\bibliography{Draft}
\bibliographystyle{unsrt}

\end{document}